\documentclass[aps,prb,twocolumn,superscriptaddress,floatfix,10pt]{revtex4-1}

\usepackage{graphicx}
\usepackage{amsmath}
\usepackage{amssymb}
\usepackage[bookmarksnumbered=true,hypertexnames=false,colorlinks=true,linkcolor=blue,urlcolor=blue,citecolor=blue,anchorcolor=green]{hyperref}

\newcommand{\ket}[1]{\mbox{$| #1 \rangle$}}
\newcommand{\braket}[2]{\mbox{$\langle #1 | #2 \rangle$}}

\begin{document}

\title{Exploring Gamification in Quantum Computing: The Qubit Factory}
\author{Glen Evenbly}
\affiliation{AWS Center for Quantum Computing, Pasadena, CA 91125, USA}
\email{evenbly@amazon.com}
\date{\today}

\begin{abstract}
Gamification of quantum theory can provide new inroads into the subject: by allowing users to experience simulated worlds that manifest obvious quantum behaviors they can potentially build intuition for quantum phenomena. The Qubit Factory\cite{TQF} (TQF) [\href{www.qubitfactory.io}{www.qubitfactory.io}] is an engineering-style puzzle game based on a gamified quantum circuit simulator that is designed to provide an introduction to qubits and quantum computing, while being approachable to those with no prior background in the area. It introduces an intuitive visual language for representing quantum states, gates and circuits, further enhanced by animations to aid in visualization. The Qubit Factory presents a hierarchy of increasingly difficult tasks for the user to solve, where each task requires the user to construct and run an appropriate classical/quantum circuit built from a small selection of components. Earlier tasks cover the fundamentals of qubits, quantum gates, superpositions and entanglement. Later tasks cover important quantum algorithms and protocols including superdense coding, quantum teleportation, entanglement distillation, classical and quantum error correction, state tomography, the Bernstein-Vazirani algorithm, quantum repeaters and more. 
\end{abstract}

\maketitle

\section{Introduction}
Quantum mechanics is a notoriously challenging subject to grasp. A significant part of this difficultly stems from the fact that individuals do not have direct experience in observing and manipulating the quantum world in the same way that they do with the classical world, thus are unable to rely on intuition to build understanding. Not only is intuition about the quantum world lacking but, more-so, quantum behavior often runs \emph{directly counter} to existing intuition, working against ones' preconceived notions of how the world operates. For instance, even something as deeply ingrained in the human experience as \emph{object permanence} is not quite true in the quantum world, where objects \emph{do} behave differently depending on whether or not they are observed.

Traditionally quantum mechanics has been introduced to students at the tertiary level as part of a degree in physics, and following a rather extensive prerequisite mathematical background. While this may have been adequate in a time when quantum theory was a topic largely confined to the academic sector, we have now entered an era in which quantum technologies, such as quantum computing\cite{{Nielsen}}, are expected to have a significant impact on society in the not too distant future\cite{Impact1, Impact2, Impact3, Impact4, Impact5} and have thus expanded into the industrial and commercial sectors at an explosive rate\cite{Growth1, Growth2}. With this rapid expansion comes a critical need for a corresponding expansion in the quantum literate workforce\cite{Workforce1, Workforce2, Workforce3, Workforce4}, with many proposals to begin the quantum education of students at a younger age\cite{Early1, Early2, Early3, Early4}. There is also a growing need to educate the broader public in quantum technology in order for stakeholders’, investors and policy-makers to make prudent decisions in this area.

Unfortunately, it often remains a significant challenge for the layperson to gain a foothold in quantum theory, whereby they can learn key concepts and ideas without necessarily going through the rigorous mathematical development needed to gain a more technically complete understanding. The development of digital learning resources\cite{OnlineResource1, OnlineResource2}, and gamified quantum resources\cite{Games1, Games2, Games3, Games4, Games5, Games6, Games7, Games8} in particular, represents a promising avenue for improving the quantum literacy of the broader public. Quantum games, by providing simulated worlds that exhibit quantum phenomena overtly, can potentially allow users to build intuition about quantum behavior from their experiences within these worlds\cite{GameTheory1, GameTheory2, GameTheory3, GameTheory4, GameTheory5}. As postured by Preskill\cite{Impact3}, “Perhaps kids who grow up playing quantum games will acquire a visceral understanding of quantum phenomena that our generation lacks”.

The Qubit Factory\cite{TQF} (TQF) is a browser-based application that serves to gamify quantum circuits and quantum computation, with a focus on differentiating quantum computing from classical computing. Intended to be accessible to those without a prior education in quantum theory, TQF challenges users to construct classical/quantum circuits in order to manipulate bits and qubits to solve computational tasks. Users progress through TQF in series of increasingly challenging levels, each aimed to illustrate a key principle of classical or quantum computation. At the start of every level, the user is presented with a variety of logical components to build with and must use their own creativity to plan and construct a solution to the specified computational task. Once their construction is complete the user can power-on their creation to watch as the computation is enacted in real-time, with animations designed to aid in the visualization and understanding of quantum processes. 

This manuscript is intended to serve as a reference guide for TQF, providing an overview of its content, functioning, and gameplay mechanics. On a broader level, we also explore the challenges inherent to the gamification of quantum theory in general, such as in representing quantum states and processes visually. We discuss how TQF attempts to resolve these issues and compare against the methods employed by previous quantum games. The manuscript is organized as follows. In Sec. \ref{sect:Premise} the premise, content and basic gameplay loop of TQF is discussed, while Sec. \ref{sect:Physics} describes the underlying physics engine. The visual designs used to represent quantum states and quantum circuits are introduced in Secs. \ref{sect:State} and \ref{sect:Circuit} respectively. In Sec. \ref{sect:Design} we elucidate the key design goals of TQF, while in Sec. \ref{sect:Level} we provide examples of how these design goals are executed within specific levels. Finally, in Sec. \ref{sect:Gamification}, we conclude with a broader discussion on the gamification of quantum mechanics and the accompanying educational benefits.

\section{Premise and Overview} \label{sect:Premise}
The Qubit Factory is intended to introduce users to qubits and quantum gates, before then exploring some foundational aspects of quantum circuits, quantum computation and quantum algorithms. Our ultimate goal is to allow users to learn and understand some of the key differences between classical and quantum computation; to this end TQF also contains substantial content related to classical computation including tasks involving bit manipulation using classical (i.e. Boolean) logic gates. Note that our focus is on theoretical aspects of quantum computing; we do not to cover topics related to the engineering challenges\cite{Engineer1, Engineer2, Engineer3} of building an actual quantum computer (e.g. like noise and decoherence), which have been the focus of other quantum games\cite{Games7}, although TQF does cover some related theory topics like quantum error correction\cite{Error1, Error2}. Similarly, we describe qubits entirely as abstract objects rather than attempting to provide any concrete realization (e.g. as ions, photons, electronic spin or other). 

In terms of gameplay, TQF takes influence from many other programming/construction style puzzle games including The Incredible Machine, titles from developer Zachtronics\cite{Zach1} (e.g. SpaceChem, TIS-100, Opus Magnum, Shenzen I/O), Human Resource Machine\cite{Zach2} and Prime Mover\cite{Zach3}. Similar to these titles, progress in TQF is divided into a series of levels, where each level the player is given some inputs (typically streams of bits and/or qubits) and a logical task to perform on these inputs in order the generate the required output(/s). An example of such a logical task could be to output the pairwise logical AND between the bits from two inputs streams. In general tasks are designed either to showcase a particular gate or operation, or to highlight a key concept from classical or quantum computation.

\begin{figure} [!t] 
\begin{center}
\includegraphics[width=8.5cm]{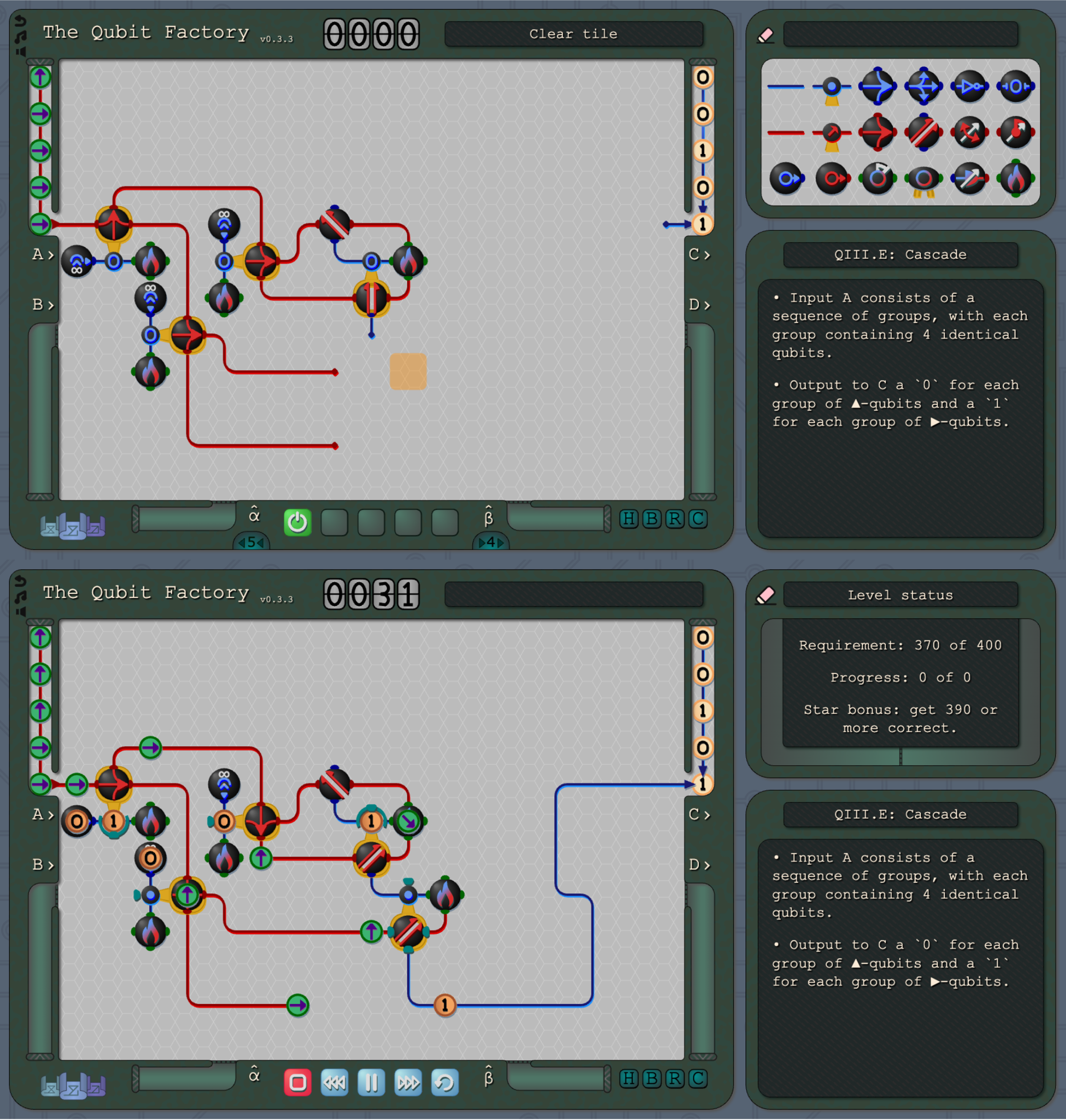}
\caption{(Top) Screenshot showing a factory midway through the \emph{construction phase}, where players design and construct circuits from the selection of components shown in the top-right window. (Bottom) Screenshot showing a factory midway through the \emph{simulation phase}, where bits and qubits move through the circuit and interact with gates in real-time. Users can speed-up, slow-down, pause and reverse the simulation via the five control buttons along the bottom of the main console.}
\label{fig:qf_construct}
\end{center}
\end{figure}

Each level begins in the \emph{construction phase}, where players design and create their factory by placing wires and gates from the set of available components, see also Fig. \ref{fig:qf_construct}. In many levels the factory floor starts blank such that users have complete freedom in how they approach the given task and they must rely entirely on their own creativity to design a valid solution. Some levels, however, are more structured and begin with existing fixed components or barriers that must be incorporated into the design. At any stage of construction users can power on their factory to enter the \emph{simulation phase}. In the simulation phase bits and qubits move through the factory, generally moving one tile per tick of the factory clock, and interact with gates and components. The speed of the clock can be sped up or slowed down, and it can also be paused or reversed. Any quantum state present can also by examined via the state analyzer as discussed further in Sec. \ref{sect:State}. These options allow users to easily assess the functionality of their current factory design. The simulation proceeds either until (i) the user chooses to halt the simulation and return to the construction phase, (ii) the level victory condition is satisfied (iii) the level victory condition is failed, (iv) or factory clock reaches 10,000 ticks (which results in automatic failure).

In most levels it would not be expected for users to reach a viable solution on their first attempt; the intended gameplay loop is for users to create a trial solution, test it via simulation, then alter the trial solution to correct any observed errors or deficiencies. In more difficult levels this process of designing, simulating, then re-designing will likely be repeated many times over until a viable solution is reached. Each level in TQF also includes an optional bonus objective designed to provide a more difficult challenge for advanced users that, when satisfied, results in the award of a coveted `bonus star' for the level. 

\section{Physics Engine} \label{sect:Physics}
Given that TQF is intended to provide \emph{authentic} knowledge about quantum computing it is important that the relevant physics is portrayed accurately. To this end TQF encompasses a quantum circuit simulator, similar to other browser-based simulators such as Quirk\cite{GUI1} or IBM Quantum Composer\cite{GUI2}, that accurately models quantum processes including unitary gates, entangled states, state measurements and wavefunction collapse.

However, whilst the simulator is accurate, it is not a \emph{complete} quantum circuit simulator; several restrictions have been artificially imposed with the goal of simplifying the user experience. One such limitation is a restriction that at most 6 qubits can be entangled together. The restriction is largely for pragmatic reasons; beyond this limit it becomes impractical to display the amplitudes of the entangled state. In practice this bound could be easily extended to at least 12 qubits, requiring 4096 parameters per state, without significantly impacting the overall memory usage or performance of the application. Currently there is no in-game reason to extend this bound as all levels can be completed without the need to entangle more than 4 qubits together at any one time. In-game, if one attempts to create an entangled state of more than 6 qubits then then the least entangled qubits from the group will spontaneously `decohere' (i.e. realized as a projective measurement) until only 6 qubits remain entangled. Another restriction imposed on the simulator is that only real-valued wavefunctions can be created. By removing the need for complex phase information (beyond +ve/-ve signs) this restriction greatly improves the ease with which quantum states can be visualized and understood by allowing a single qubit state to be represented as an arrow on a 2D disk (rather than as an arrow in the 3D Bloch sphere). These aspects of quantum state representations are discussed further in Sec. \ref{sect:State}. Inherent to the restriction of real-valued wavefunctions is a corresponding limitation on the set of allowed quantum gates, which we consider shortly.

\begin{figure} [!t!h] 
\begin{center}
\includegraphics[width=7.0cm]{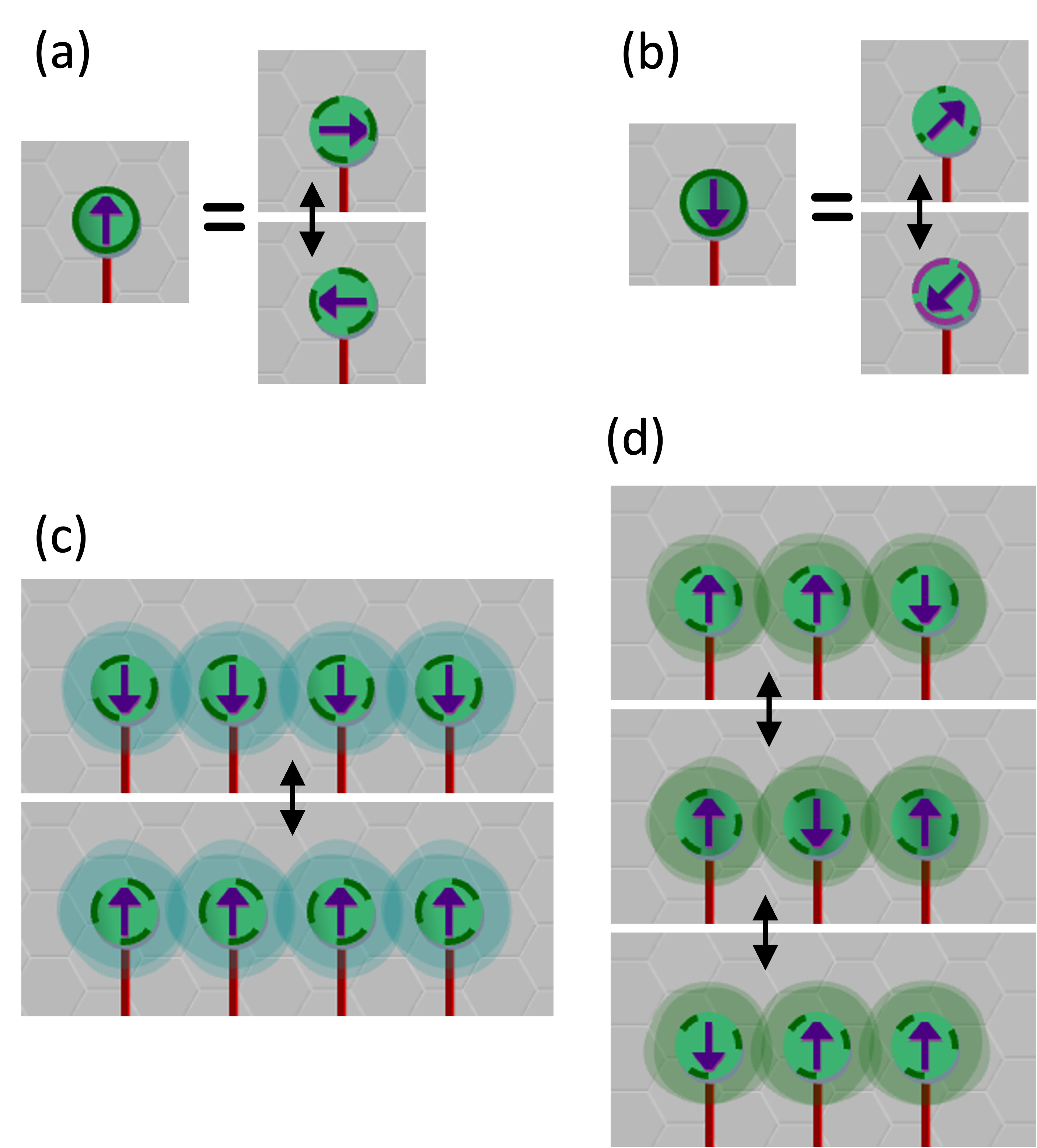}
\caption{(a) The state \ket{\uparrow} can equivalently be represented as an even superposition of states \ket{\gets} and \ket{\to}, see Eq. \ref{eq:super1}. In-game, the superposition is indicated by the qubit `blinking' back and forth between the configurations shown. (b) The state $\ket{\downarrow}$ can equivalently be represented as a superposition of states $\ket{\nearrow}$ and $\ket{\swarrow}$, see Eq. \ref{eq:super2}. Notice that the dash-length of outline denotes the magnitude of each wavefunction amplitude, while the outline color denotes its sign (with green for positive and purple for negative). (c) A GHZ state on four qubits is depicted as `blinking' between the two basis states of non-zero amplitude, see also Eq. \ref{eq:ghz}. The translucent cloud surrounding each of the qubits matches in color to indicate that the four qubits are part of the same entangled state. (d) The W-state, see Eq. \ref{eq:w}, is depicted as blinking between the three basis states of non-zero amplitude.}
\label{fig:qf_blink}
\end{center}
\end{figure}

Rather than presenting to the player an exhaustive set of classical/quantum gates and components to build with, our philosophy in designing TQF was to only include the \emph{minimal set} of gates necessary to construct interesting challenges and to showcase the intended concepts. On the side of classical computation we include only two single-bit gates, the `inversion' and `re-zero' (or constant zero) gates, together with a \emph{control} construct that can toggle the function of a target gate dependant on an input control bit. When paired with the single bit inversion or re-zero gates, the control/target construct can form `XOR' and `AND' logic gates respectively, thus they constitute a universal gate-set for classical computation. On the side of quantum computation we include two single qubit gates, the `flip' and `rotate' gates, which both have a configurable axis that can point at angle $\theta\in (-\pi,\pi]$ in discrete increments of $\pi/8$. The flip gate acts to mirror a qubit in the $X$-$Z$ plane about the $\theta$-axis, thus can account for Pauli-Z, Pauli-X and Hadamard gates for axes $\theta=\{0,\tfrac{\pi}{2},\tfrac{\pi}{4}\}$ respectively. The rotation gate performs a $Y$-rotation of qubits through the angle $\theta$. These gates, which can also be supplemented by a quantum control, are discussed in more detail in Sec. \ref{sect:Circuit}. Also included is a single-qubit measurement component, which can output a classical bit signifying the measurement result as well as the post-measurement (i.e. projected) qubit. Given the lack of any phase-shift gate it is clear that this gate-set is not universal for quantum computation; none-the-less we advocate that the gate-set is sufficiency broad to cover a wide range of concepts, challenges and algorithms from quantum computing.

Finally, we remark that the simulation engine of TQF possesses some functionality and flexibility not present in standard classical/quantum circuit simulators. In particular, circuits in TQF are not restricted to linear flow from left-to-right; bits/qubits move along wires which can be placed in any direction and can even loop back around on themselves. This non-linear control flow allows for simple realizations of `FOR' and `WHILE' loops, facilitating the construction of relatively sophisticated algorithms in a compact space and with few gates. Additionally, in contrast to standard circuit models where each wire accounts for a single bit or qubit, wires in the TQF can simultaneously hold multiple bits or qubits, which move along wires serially. The allowance of multiple bits/qubits per wire has a variety of benefits: 
\begin{itemize}
    \item It facilitates the design challenges involving \emph{efficiency}, where the goal is to process a stream of inputs in as short of a time as possible.
    \item It aids in the understanding of the probabilistic nature of quantum mechanics, given that it can allow the user to see multiple outcomes of a process in quick succession and on the screen at the same time.
    \item It adds elements of timing and synchronization in order to provide an extra layer of depth to the gameplay.
\end{itemize}
A more subtle benefit of allowing multiple bits/qubits per wire is that it brings the gameplay closer to established titles in the popular factory-building genre (e.g. Factorio\cite{Factorio1}), where working with production lines conveying multiple objects simultaneously is common.

\begin{figure} [!t] 
\begin{center}
\includegraphics[width=6.5cm]{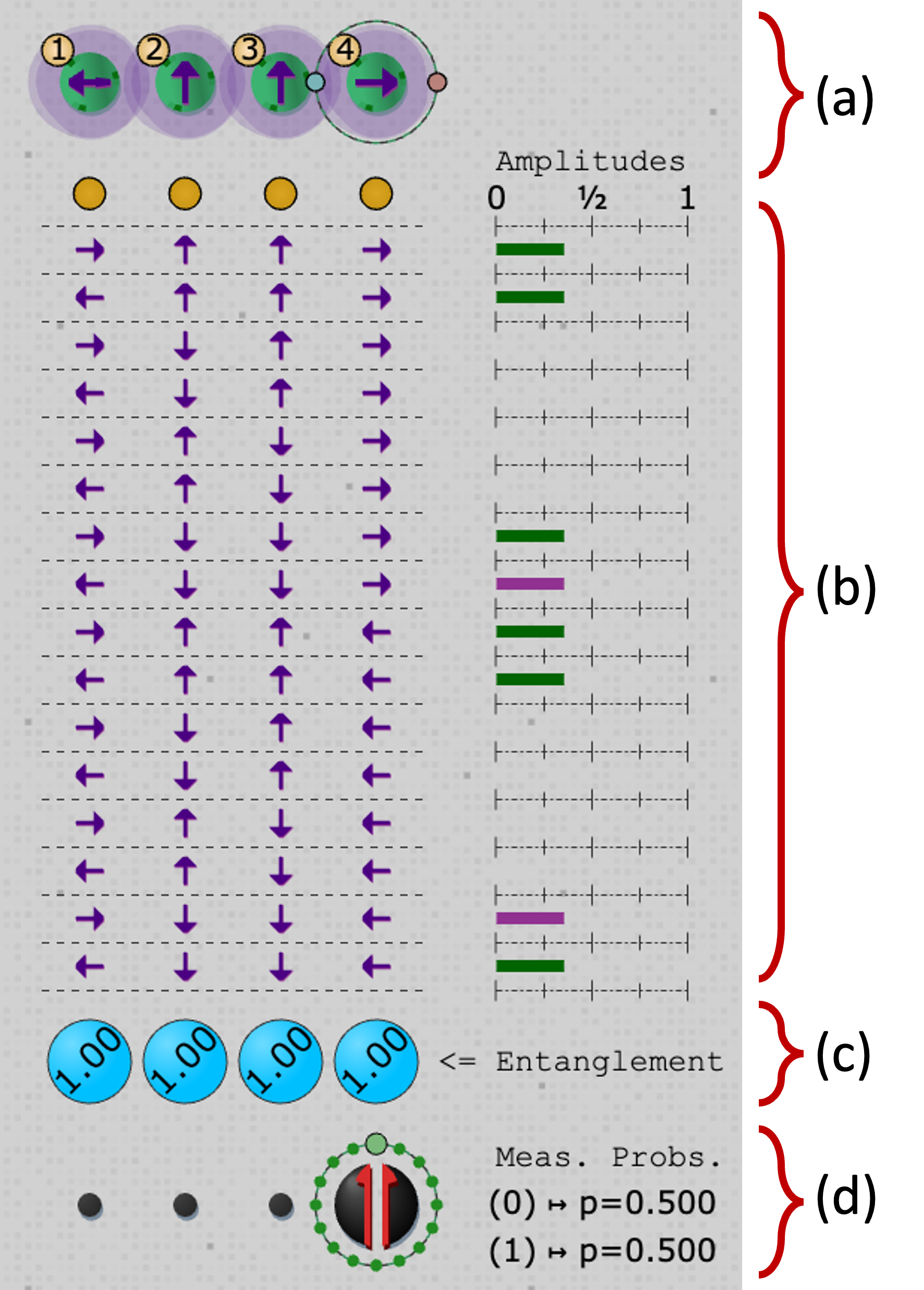}
\caption{Screen-shot of the state analyzer acting on an entangled state of 4 qubits. (a) The basis of the selected qubit can be changed by dragging its control points. (b) Display of the $2^4 = 16$ basis states together with the wavefunction amplitudes in each, where green/purple coloration represent positive/negative signs respectively. (c) The entanglement entropy of each individual qubit, computed as per Eq. \ref{eq:ent}. (d) The measurement outcome probabilities of the selected qubit under the measurement shown, which the user can adjust to any angle.}
\label{fig:qf_analyze}
\end{center}
\end{figure}

\section{Quantum State Representations} \label{sect:State}
A major challenge in any attempt to gamify quantum mechanics comes with the need to represent quantum states visually; preferably in a way that can make intuitive sense to users. In this section we discuss visual language used in TQF to represent quantum states, beginning with (unentangled) single-qubit pure states. A particular illuminating way to represent a pure state of a qubit is geometrically as a point on the surface of the Bloch sphere. Given that TQF only considers real-valued wavefunctions if follows that the allowed qubit states can be represented as a point on the surface of a disk by fixing the azimuthal angle $\varphi=0$ in the Bloch sphere\cite{Bloch}. This is the representation we employ in TQF; the state $\ket{\psi(\theta)}$ at polar angle $\theta\in (-\pi,\pi]$ on the Bloch sphere,
\begin{equation}
\ket{\psi(\theta)} \equiv \cos\left(\tfrac{\theta}{2}\right) \ket{\uparrow} + \sin\left(\tfrac{\theta}{2}\right) \ket{\downarrow}, \label{eq:bloch}
\end{equation}
is represented in-game by a disk with an arrow pointing in direction $\theta$, see Fig. \ref{fig:qf_blink}. However, players are also allowed to alter the basis of an qubit as they please; thus a state $\ket{\psi(\theta)}$ could potentially be represented as a superposition of any pair of orthogonal states. For instance, the player could choose to represent the $\ket{\uparrow}$ state as a superposition of $\ket{\to}$ and $\ket{\gets}$ states (or equivalently the $\ket{\psi(\tfrac{\pi}{2})}$ and $\ket{\psi(-\tfrac{\pi}{2})}$ states respectively),
\begin{equation}
\ket{\uparrow} = \left( \tfrac{1}{\sqrt{2}} \right) \ket{\to} + \left( \tfrac{1}{\sqrt{2}} \right) \ket{\gets}. \label{eq:super1}
\end{equation}
In-game this superposition would be represented by the qubit `blinking' between each of the basis states, where the magnitude of each amplitude in the superposition is indicated both by the amount of time spent in the basis state and also by the length of the dashes along the edge of the qubit disk, see also Fig. \ref{fig:qf_blink}(a). Let us consider a second example, where we represent state $\ket{\downarrow}$ in the basis states at rotations $\theta = \tfrac{\pi}{4}$ and $\theta = \tfrac{-3\pi}{2}$,
\begin{equation}
\ket{\downarrow} = \sin \left( \tfrac{\pi}{8} \right) \ket{\nearrow} - \sin \left( \tfrac{3 \pi}{8} \right) \ket{\swarrow }, \label{eq:super2}
\end{equation}
see also Fig. \ref{fig:qf_blink}(b). In this case the $\ket{\swarrow }$ component is represented in-game as a disk with a purple border (as opposed to the usual green border) to denote that the sign of the amplitude is negative.

Entangled states of multiple qubits are represented similarly: consider for instance a GHZ state on a set of four qubits,
\begin{equation}
\ket{\textrm{GHZ}} = \left( \tfrac{1}{\sqrt{2}} \right) \ket{\uparrow \uparrow \uparrow \uparrow}  + \left( \tfrac{1}{\sqrt{2}} \right) \ket{\downarrow \downarrow \downarrow \downarrow}. \label{eq:ghz}
\end{equation}
In-game the GHZ state is represented by having the four qubits `blink' back and forth between the $\ket{\uparrow \uparrow \uparrow \uparrow}$ and $\ket{\downarrow \downarrow \downarrow \downarrow}$ states, as shown in Fig. \ref{fig:qf_blink}(c). However, an additional visual cue is needed to signify which qubits are entangled with each other; we address this issue by surrounding any qubit that is part of an entangled state with a translucent cloud that matches in color between qubits that are part of the same entangled state. The size of the cloud surrounding each qubit also proportionate to its single qubit entanglement entropy,
\begin{equation}
\mathcal{S}(\rho) = \textrm{Tr} \left( \rho_A \log \rho_A \right), \label{eq:ent}
\end{equation}
where $\rho_A$ is the density matrix corresponding to the qubit, which provides users with a rough visual indication of how entangled each qubit is. As another example we consider a W-state on three qubits,
\begin{equation}
\ket{\textrm{W}} = \left( \tfrac{1}{\sqrt{3}} \right) \ket{\uparrow \downarrow \downarrow}  + \left( \tfrac{1}{\sqrt{3}} \right) \ket{\downarrow \uparrow \downarrow } + \left( \tfrac{1}{\sqrt{3}} \right) \ket{\downarrow  \downarrow \uparrow}, \label{eq:w}
\end{equation}
which is represented in-game by the qubits `blinking-through' the three basis states with non-zero weight, as shown in Fig. \ref{fig:qf_blink}(d). 

While the visual representation of superpositions and entangled states as `blinking-through' configurations may be sufficient to convey rough information on simple states, such as Bell pairs or GHZ states, it is undoubtedly inadequate to describe more complicated states. To address this deficiency, TQF incorporates a \emph{state analyzer} that appears whenever a qubit is selected (either during the construction or the simulation phase), whose purpose is to provide additional and more precise information, see Fig. \ref{fig:qf_analyze}. If the selected qubit is part of a $N$-qubit entangled state then the state analyzer will display the $2^N$ wavefunction amplitudes (or the $16$ largest magnitude amplitudes if $N>4$). The state analyzer allows users to change the basis of any qubit by dragging its control points, thus to easily explore how the state amplitudes appear under a different choice of basis. Additionally, the analyzer also displays the single qubit entanglement entropies as given in Eq. \ref{eq:ent} as well as the outcome probabilities for any single qubit measurement. A valuable function that the analyzer provides is that it allows users to \emph{predict} the resulting wavefunction if one qubit from an entangled state were to be measured along a specific axis; this functionality is necessary to solve some of the more difficult levels in TQF, such as those involving entanglement distillation\cite{Algo1,Algo2}.

\begin{figure} [!t] 
\begin{center}
\includegraphics[width=5.5cm]{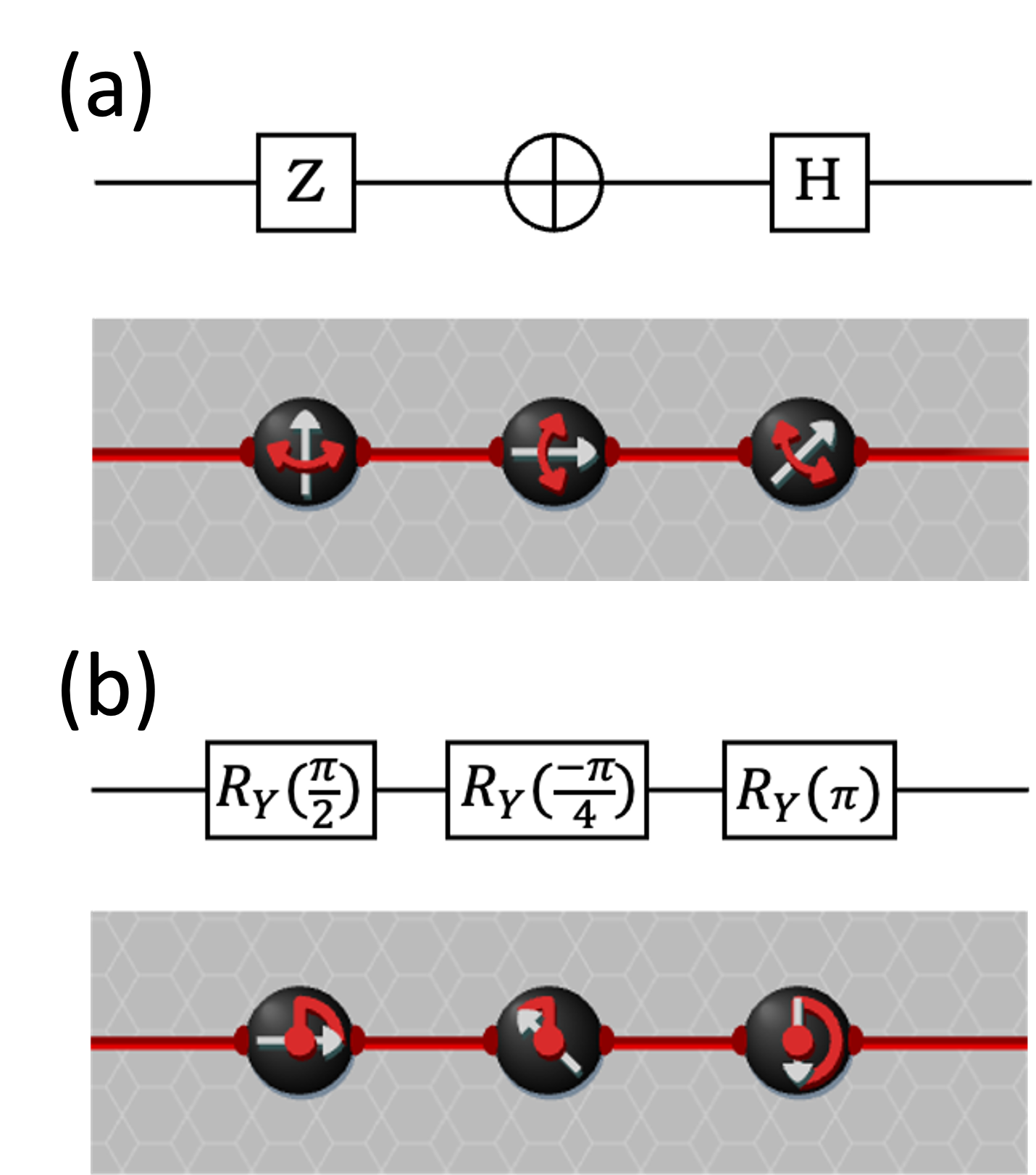}
\caption{(a) Conventional circuit notation for Pauli-Z, Pauli-X, and Hadamard (H) gates, together with their in-game depictions where they are represented via the `flip' gate set at angles $\theta = \{0, \tfrac{\pi}{2}, \tfrac{\pi}{4} \}$ respectively. (b) Conventional notation for $Y$-rotations through angles $\theta = \{\tfrac{\pi}{2}, -\tfrac{\pi}{4}, \pi \}$ respectively, together with their in-game depictions where the equivalent gates are represented via the `rotate' gate set to the angles $\theta$.}
\label{fig:gate1}
\end{center}
\end{figure}

\section{Circuit Representations} \label{sect:Circuit}
In this section we discuss visual language used in TQF to represent quantum gates/components and their actions on qubits. Although there exists widely established notation for quantum circuits\cite{Nielsen}, we have chosen to deviate from the established norm and to instead introduce our own custom representations. There are several reasons for this choice:
\begin{itemize}
    \item Ideally we want users to be able to infer the function of a gate from its design alone, rather than having to memorize the function of each gate (as is the case in the established notation where gates are often represented by a single letter, e.g. the H gate). This is intended to remove a barrier-to-entry for newcomers to the space of quantum circuits, as well as to lower the learning curve required for users to design their own circuits. This visual design is further enhanced through the in-game animations, which can provide additional cues to the user on the function of each gate. 
    \item We wish to provide gate/components that possess more flexibility than the standard notation allows for. For instance, the standard circuit notation for a measurement assumes that it is taken in the $Z$-basis, but in TQF we instead use a design that can depict a measurement taken in any specified basis.
\end{itemize}
It should be noted that Quantum Flytrap\cite{Games8} also employed a custom representation of components (e.g. a flytrap to represent a photon detectors) for similar reasons as outlined above.

\begin{figure} [!t] 
\begin{center}
\includegraphics[width=5.5cm]{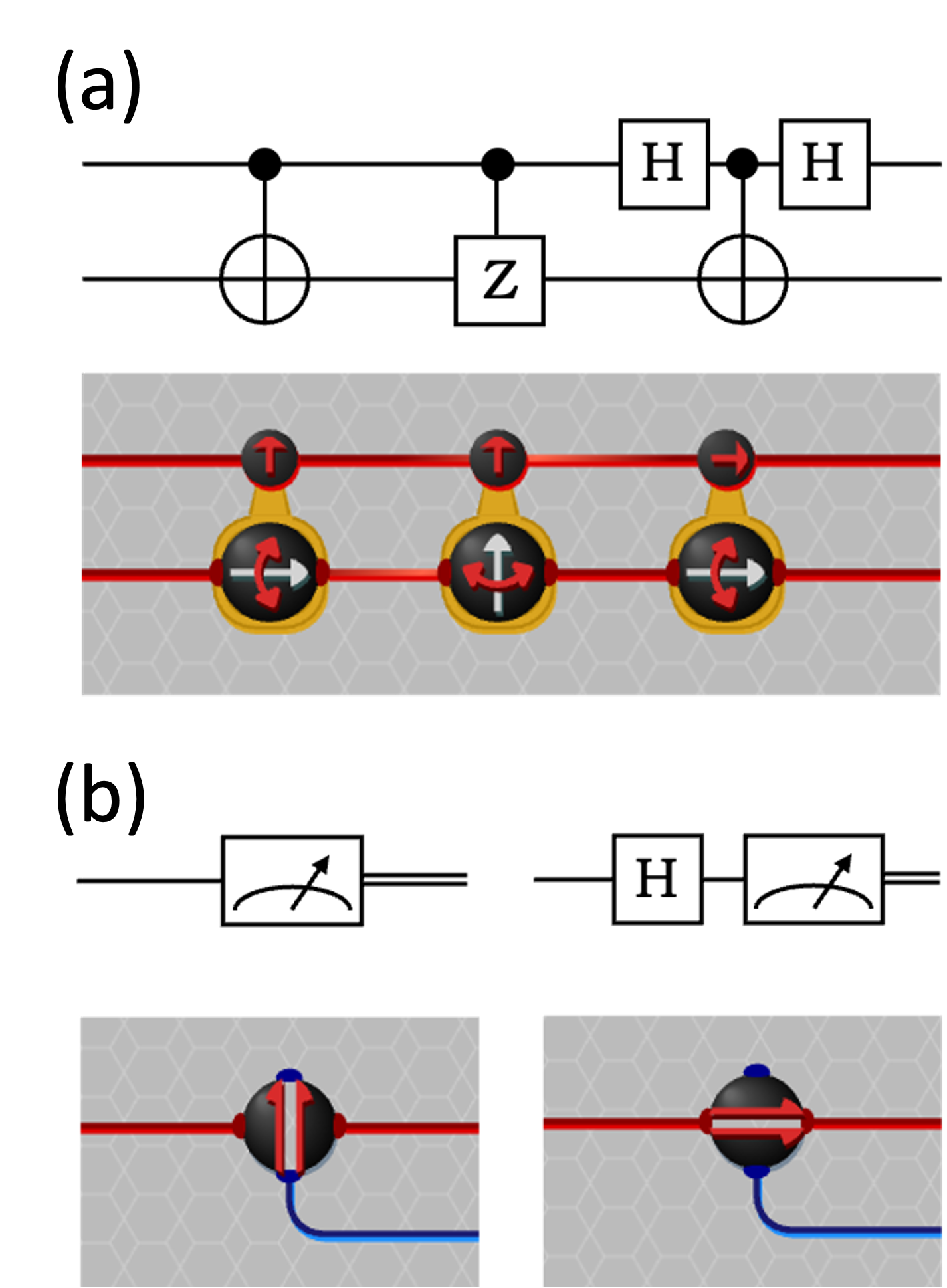}
\caption{(a) Conventional circuit notation for controlled-not (CNOT), controlled-Z (CZ), and a CNOT where the control has been sandwiched between a pair of Hadamard gates (and is thus equivalent to an $X$-axis control), together with in-game depictions of equivalent gates. In TQF the axis in which the control acts can be set at any angle $\phi$; thus the $X$-axis control can be realized without the need for Hadamard gates. (b) Conventional notation for $Z$-axis and $X$-axis measurements (the latter of which is implemented as the Hadamard gate followed by an $Z$-axis measurement), each of which produce as output a classical bit as denoted by the double-line. Below are in-game depictions of equivalent measurements in TQF, where the axis of the measurement is configurable via the orientation of the slit, which can return both a classical bit (via the blue wire) as well as the post-measurement qubit (via the red wire) as output.}
\label{fig:gate2}
\end{center}
\end{figure}

\begin{figure} [!t] 
\begin{center}
\includegraphics[width=5.5cm]{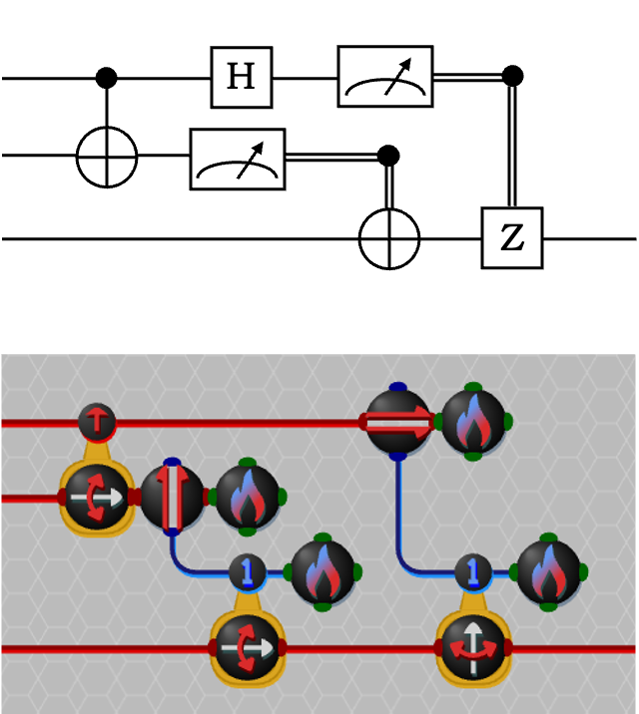}
\caption{(Top) Conventional circuit notation for an implementation of the quantum teleportation protocol\cite{Algo3}. (Bottom) An in-game depiction on an equivalent circuit. This implementation circumvents the need for a Hadamard gate, since the $X$-axis measurement can be taken directly, but requires additional `incinerators` to remove bits/qubits once they are no longer needed.}
\label{fig:teleport}
\end{center}
\end{figure}

One of the single qubit gates featured in TQF is the \emph{flip} gate $F(\theta)$, which acts to flip a qubit about some axis $\theta \in (-\pi,\pi]$ and has the matrix representation
\begin{equation}
F(\theta) = 
\begin{bmatrix}
\cos(\theta) & \sin(\theta) \\ 
\sin(\theta) & -\cos(\theta)
\end{bmatrix}.
\end{equation}
The flip gate is represented in-game as a black tile with a white arrow pointing in direction $\theta$, coupled with a red doubled-headed arrow intended to signify reflection about the $\theta$-axis, see Fig. \ref{fig:gate1}(a). This gate can account for Pauli-Z, Pauli-X and Hadamard gates with rotations $\theta = \{0, \tfrac{\pi}{2}, \tfrac{\pi}{4}\}$ respectively. The visual design is intended to be intuitive such that, for instance, a user without background quantum knowledge could reasonably guess that the $F(\pi/4)$ gate would reflect a $\ket{\uparrow}$ qubit into a $\ket{\to}$ qubit. This intuition is further augmented by the in-game animations, which show qubits flipping about the gate axis when the gate is applied.

The second single gate featured in TQF is the \emph{rotate} gate $R(\theta)$ which acts to rotate qubits about the $Y$-axis through the rotation angle $\theta \in (-\pi,\pi]$ and has the matrix representation
\begin{equation}
R(\theta) = 
\begin{bmatrix}
\cos(\theta/2) & -\sin(\theta/2) \\ 
\sin(\theta/2) & \cos(\theta/2)
\end{bmatrix}.
\end{equation}
The rotate gate is represented in-game via a black tile with a white arrow pointing in direction $\theta$, coupled with a red circle intended to denote the rotational pivot and a red line denoting the extent of rotation, see Fig. \ref{fig:gate1}(b). Once again, this visual design is intended to allow users to infer that it acts to rotate qubits, which is also shown via the in-game animations whenever the gate acts on a qubit. 

The third component that we consider is the quantum control $C_U(\phi)$, which acts in conjunction with a secondary unitary gate $U$, which is required to be either a flip $F(\theta)$ or rotate $R(\theta)$ gate. The transformation of the control/target qubit pair can be represented by a $4\times 4$ matrix,
\begin{equation}
C_{U}(\phi) = 
\begin{bmatrix}
\mathbb{I} - \sin^2(\phi/2) A & \sin(\phi) A \\ 
\sin(\phi) A & U + \sin^2(\phi/2) A 
\end{bmatrix},
\end{equation}
where we have defined $A \equiv (\mathbb{I}-U)$. The angle $\phi\in (-\pi,\pi]$, which sets the basis in which the control acts, can be configured by users during the construction phase. Setting $\phi=0$ yields the usual quantum control, while setting $\phi=\pi$ yields an anti-control, and setting $\phi=\tfrac{\pi}{2}$ yields an $X$-axis control. It follows that $C_{F(\pi/2)}(0)$ reproduces the CNOT quantum gate,
\begin{equation}
C_{F(\pi/2)}(0) = 
\begin{bmatrix}
1 & 0 & 0 & 0 \\
0 & 1 & 0 & 0 \\
0 & 0 & 0 & 1 \\
0 & 0 & 1 & 0 
\end{bmatrix}, 
\end{equation}
and $C_{F(0)}(0)$ reproduces the CZ quantum gate, 
\begin{equation}
C_{F(0)}(0) = 
\begin{bmatrix}
1 & 0 & 0 & 0 \\
0 & 1 & 0 & 0 \\
0 & 0 & 1 & 0 \\
0 & 0 & 0 & -1 
\end{bmatrix}.
\end{equation}
The control $C_U(\phi)$ is represented in-game by a small black disk with a red arrow pointing in the direction $\phi$, see also Fig. \ref{fig:gate2}(a). Additionally, a yellow ring extends from the control to an adjacent tile, upon which the target gate $U$ can be placed. 

The fourth component that we consider is the single qubit measurement $M(\theta)$ oriented at angle $\theta\in (-\pi,\pi]$. Let us define the eigenstate $\ket{\psi^+}$ that is \emph{aligned} with $M(\theta)$ as 
\begin{equation}
\ket{\psi^+} = \ket{\psi(\theta)},
\end{equation}
and the eigenstate $\ket{\psi^-}$ that is \emph{anti-aligned} with $M(\theta)$ as 
\begin{equation}
\ket{\psi^-} = 
\begin{cases}
 & \ket{\psi(\theta-\pi)} \text{\ if \ } \theta \ge 0 \\ 
 & \ket{\psi(\theta+\pi)} \text{\ if \ } \theta < 0
\end{cases},
\end{equation}
with the states $\ket{\psi(\theta)}$ as defined in Eq. \ref{eq:bloch}. This component $M(\theta)$ enacts a project measurement of a qubit: an initial state $\ket{\psi}$ is projected onto $\ket{\psi^+}$ with probability $p^+=\left|{\braket{\psi^+}{\psi}}\right|^2$ and onto $\ket{\psi^-}$ with probability $p^-=\left|{\braket{\psi^+}{\psi}}\right|^2$. If the qubit undergoing measurement in not part of an entangled state then the probability of projecting onto the aligned/anti-aligned eigenstates evaluates to $p^+ = \cos^2(\sigma)$ and $p^- = \sin^2(\sigma)$ respectively, where $\sigma$ is the angular difference between the orientation of the measurement and the polar angle of the qubit on the Bloch sphere. The measurement component $M(\theta)$ is represented in-game as a black tile with a slit and a red arrow pointing at angle $\theta$, see Fig. \ref{fig:gate2}(b). By default the measurement component will only output the measured (i.e. post-projection) qubit; however if one or both of its lateral blue tabs are connected to blue wires then it will also output classical bit(s) signifying the measurement result, with 0/1 bits for an aligned/anti-aligned result respectively.

Finally, we remark that in addition to the classical gates (which we do not detail explicitly since they follow a standard implementation), TQF also contains a selection of miscellaneous gates. These include `combiners' for merging wires together, `creation' gates for creating bits/qubits, `incinerators' for destroying bits/qubits, and gates related to circuit timing (`sync' and `delay'). Many quantum circuits can straight-forwardly be implemented in TQF with few changes, such as the example of the quantum teleportation\cite{Algo3} protocol show in Fig. \ref{fig:teleport}. 

\begin{figure} [!t] 
\begin{center}
\includegraphics[width=6.0cm]{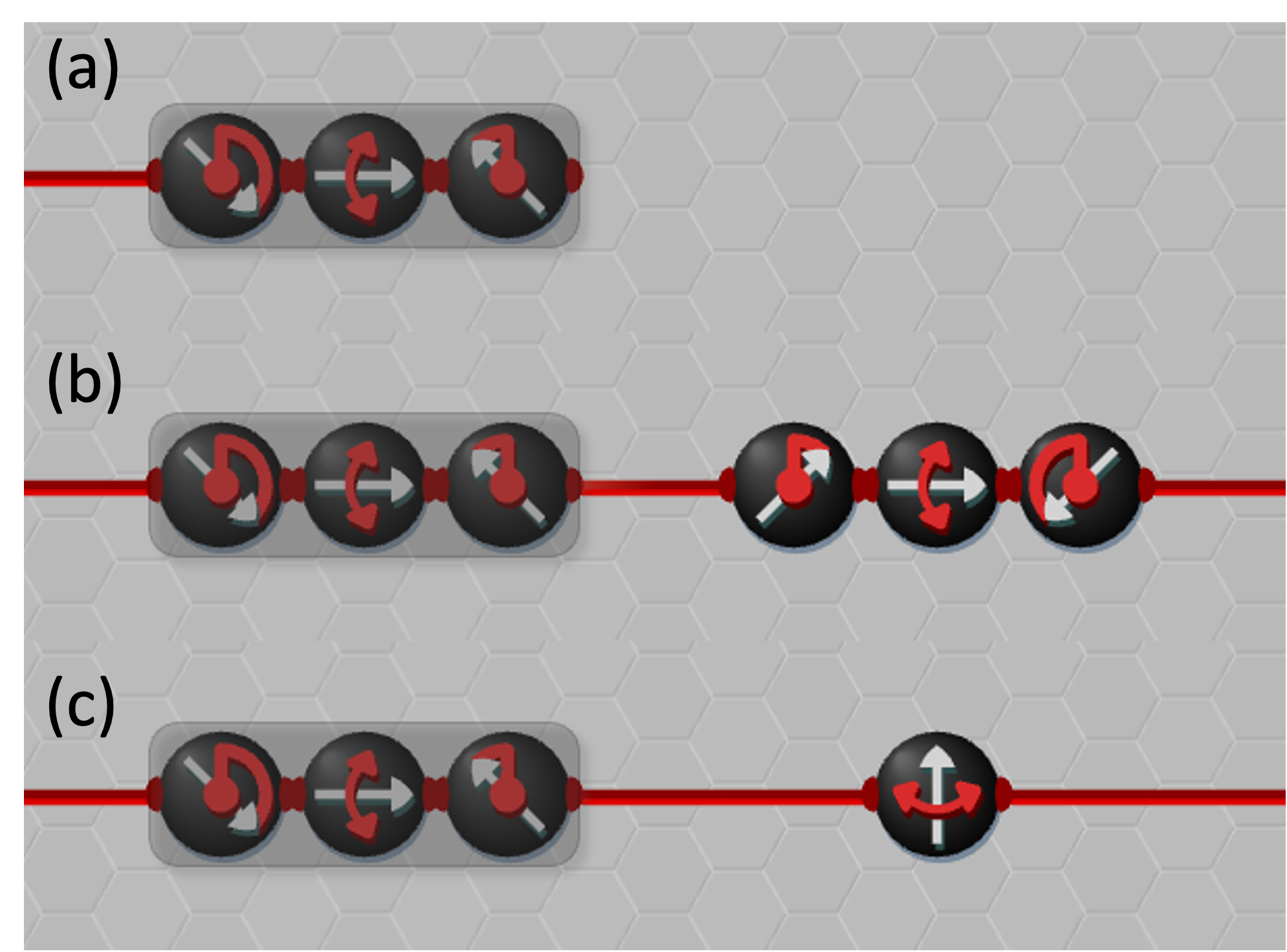}
\caption{(a) The task from level QI.A requires the user to place gates in order to `undo' the action of a sequence of pre-set rotation/flip gates. (b) An example solution to the task, where the action of each of the pre-set gates is undone singularly (and in reverse order). (c) A valid solution can also be implemented via a single flip gate.}
\label{fig:group}
\end{center}
\end{figure}

\section{Game Design} \label{sect:Design}
In the setting of a conventional tertiary education, quantum mechanics is typically taught to students who are already well-versed in the pre-requisite mathematical foundations (i.e. calculus, statistics and linear algebra), with many of the details, concepts and explanations built squarely upon these foundations. Thus, as already discussed in the introduction to this manuscript, it remains a formidable challenge to successfully teach concepts in quantum mechanics without assuming (and thus exploiting) that the learner has these mathematical foundations. However video-games, as an interactive medium, provide novel ways for users to learn and discover new ideas that go beyond conventional learning resources (i.e. textbooks and lectures). By allowing to interact with quantum systems in real time and explore through trial-and-error, it may be possible for users to build intuition and understanding from this first-hand experience\cite{GameTheory1, GameTheory2, GameTheory3, GameTheory4, GameTheory5}. In this section we outline some of TQF's key design goals and discuss how they leverage the interactive nature of games in order to convey concepts from quantum computation more effectively.

\subsection{Engage, don't tell:}
We attempt to leverage interactive nature of the games medium where-ever possible; in particular we aim to have players `discover' concepts for themselves rather than explaining them through in-game text. For instance when introducing new gates to the player, rather than directly explaining the function of the gate, the player is instead presented with a simple task designed specifically to showcase usage of that gate. This allows players to potentially discover for themselves the function of each gate through experimentation and trial-and-error, which is intended foster engagement as well as to help players build intuition. However, some more traditional resources are provided as a fall-back to prevent players from getting stuck; this includes an `employee handbook' which describes the function of each gate in detail and provides examples of usage, as well as providing a reference for other quantum concepts. Several of the more difficult levels also offer an `ex-employee journal' which aim to further contextualize the given task as well as to provide direct hints/instructions for its successful completion. Furthermore, many levels also present the user with a `historical archive' link upon completion, which links to a Wikipedia article relevant to the level they completed. These are intended to provide the interested user with the option for a more detailed and comprehensive explanation of the task that they have just explored, but without disrupting the gameplay flow for other users.

\subsection{Foster creativity:} Many of the levels in TQF are designed to be open-ended; the user if presented with a task and a set of components to build with, but the factory otherwise begins as a blank slate. In these levels there is not a single intended solution and, in fact, there may be multiple fundamentally different strategies/approaches that the user could employ to reach a viable solution. This open-ended design is intended to encourage users to experiment and to employ creativity in the design of a solution. It also adds an extra layer of engineering challenge beyond achieving a theoretical understanding of a task or concept; even if a user understands the theory underlying how a given task could be completed, the design and implementation of a factory that properly executes this theory may still require significant effort. This style of open-ended challenges is in contrast to traditional textbook-style problems which are often based on using a prescribed methodology in order to reach a unique solution to a given problem. In addition to the basic success criteria required to complete a level, each level also contains optional `bonus star' criteria designed to provide an extra challenge for advanced users. The bonus star criteria can include (i) restrictions on the number and/or type of gates that can be used, (ii) restrictions on the running-time of the factory, (iii) restriction on the amount of floor-space used, (iv) or the requirement to satisfy a higher accuracy threshold. In order to achieve the bonus star users must often streamline their designs, reduce their usage of space and/or components, and reduce the processing time taken by the factory. These optimizations are intended to reflect actual research and development undertaken by scientists and engineers in the field of quantum computing, where it is a commonplace activity to try to optimize a process or algorithm within a set of spatial and/or temporal constraints. Another way that TQF encourages creativity is by facilitating the sharing of level designs between different users: by pressing Ctrl+C within a level a user can serialize their current design to a text string that is copied to the clipboard. This string can be sent to other users (e.g. via email or text chat) who can use Ctrl+V to paste the design into their own factory.

\subsection{Slowly but surely:}
The Qubit Factory aims for a slow and gradual rate when introducing new concepts. This is especially important, not only to avoid overwhelming the user with information, but also to give users the opportunity to learn using first-hand experience and to build intuition. Similarly, we attempt a breadth-first approach to the presentation of information; rather going into detail about each concept when first introduced (as is more common in an academic setting), TQF is designed to initially offer a shallow explanation over a breadth of many different concepts, while slowly layering in depth as users progress. For instance, qubits are initially presented to the user simply as arrows on a disk, without any elaboration or further details. It is only after the users progress through several levels that the more intricate features of qubits (such as basis changes and superpositions) are introduced. 

\begin{figure} [!t] 
\begin{center}
\includegraphics[width=8.5cm]{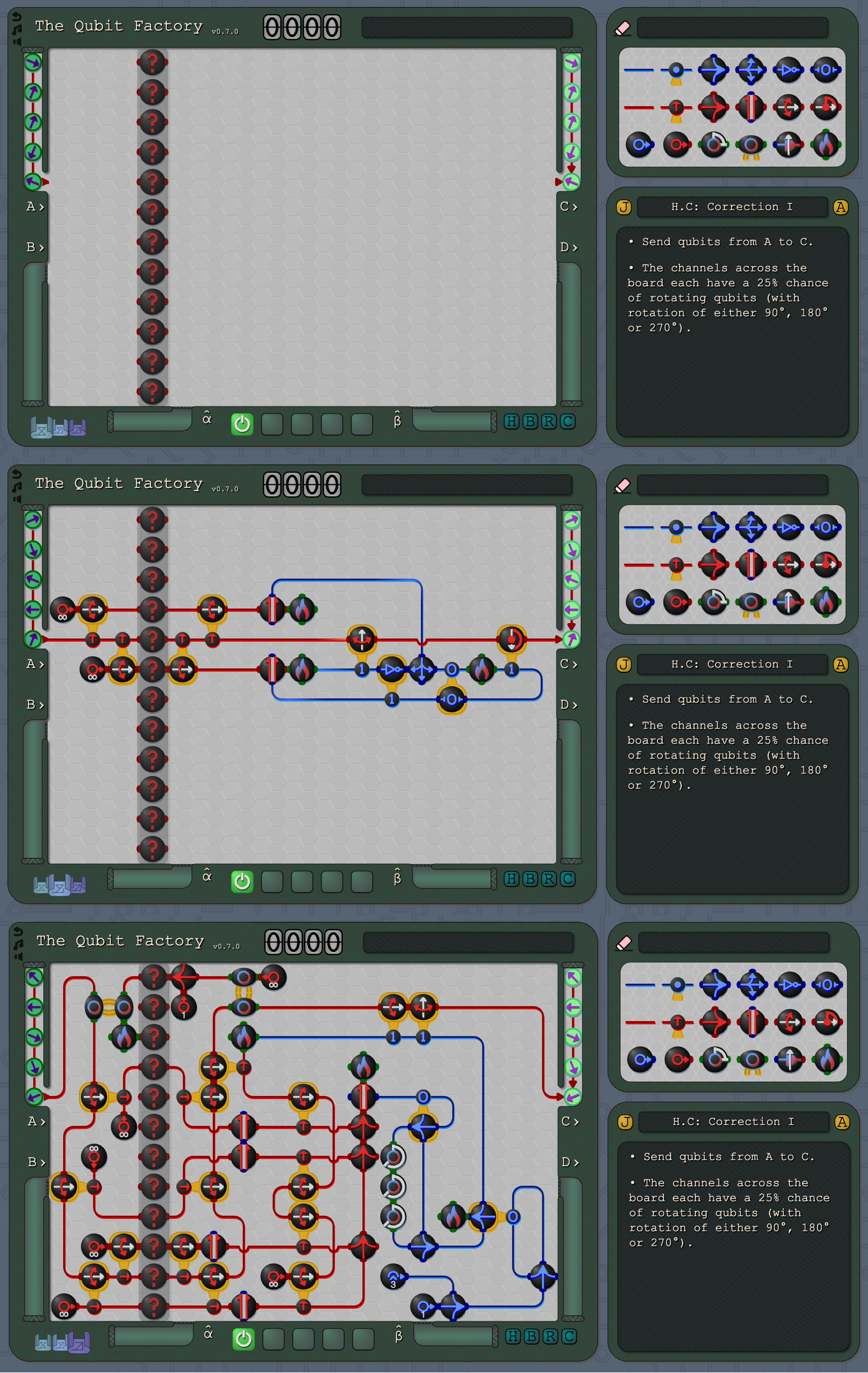}
\caption{(Top) An example level where the user is tasked with a problem in quantum error correction: users must design a factory that allows the input stream of qubits on the left to be transmitted through a noisy channel (denoted by ``?" gates) to the output on the right, while exceeding a set accuracy threshold. (Middle) An example of a solution to the level based on a bit-flip code; each data qubit is entangled with a pair of ancilla qubits prior to transmission through the noisy channel. (Bottom) An example of a more sophisticated solution designed to achieve a higher accuracy threshold (and thus to satisfy the optional `bonus star' challenge).}
\label{fig:err}
\end{center}
\end{figure}

\section{Level design:} \label{sect:Level}
\subsection{Examples:}
Let us now consider a few examples from specific levels in order to illustrate how TQF incorporates the design goals as outlined in Sec. \ref{sect:Design}.

As a first example we consider the earliest quantum-focused level presented to the user, level QI.A, where the player is presented with a sequence of pre-placed rotate/flip gates and tasked with placing additional rotate/flip gates as to `undo' the action of the existing gates, see Fig. \ref{fig:group}(a). Details of these gates are not explained to users up-front; instead it is intended that users can infer their function through observation of their action on qubits. It is hoped that players will be able to devise a solution from basic geometry alone: that any rotation can be undone by an equal rotation in the opposite direction and that any flip (or reflection) about some axis can be undone by applying the same flip again. It follows that any sequence of flips/rotates can be undone by undoing each individual gate of the sequence. However, as shown in Fig. \ref{fig:group}(b), the ordering must be reversed; since the \emph{last} action performed in the sequence in Fig. \ref{fig:group}(a) was a rotation by $\theta=-\pi/4$ it makes sense that the \emph{first} action to undo the sequence should be rotation by $\theta=+\pi/4$. Thus, upon completion of the level, users may have acquired some geometric understanding of several concepts usually understood in terms of linear algebra and matrix operators: (i) quantum gates are reversible (i.e. unitary), (ii) quantum gates don't necessarily commute, (iii) a sequence of gates can be undone by applying the inverse of the individual gates in reverse order. While the solution depicted in Fig. \ref{fig:group}(b) does complete the main level task it fails to meet the bonus star objective, which restricts to using a single gate to undo each sequence of pre-placed gates. One strategy that users could employ to satisfy this bonus objective is to simply compare the transformed qubits against the requested outputs, from which a single gate that produces these outputs can be inferred, see also Fig. \ref{fig:group}(c). The bonus objective is intended to highlight a concept usually understood in terms of group theory: (iv) a sequence of rotate/flip gates is equivalent to a single instance of some other rotate/flip gate. 

As a second example we consider levels CII.E and H.C, which are designed showcase classical and quantum error correction\cite{Error1, Error2} respectively, and to highlight the differences between them. In these levels the user must transmit a sequence of bits or qubits through a noisy channel and correct errors that occur in order to reach a prescribed accuracy threshold. Both levels begin with a blank factory, as shown Fig. \ref{fig:err}(a), such that users must determine for themselves an overall error correction strategy as well as a specific design. In the classical error correction level (CII.E), we think it likely that most users (even those without any prior knowledge in error correction) would think to duplicate the data before transmission in order to employ some kind of repetition code, as the most obvious fix to the occurrence of random errors is simply to send more than one copy of the data. Upon later reaching the quantum error correction level (H.C) users may be tempted to repeat the same repetition strategy from the classical error-correction level. However, if they attempt as such they will be hit with an immediate roadblock: that, unlike classical bits, qubits cannot cloned\cite{Algo4} so using a repetition strategy is not feasible. Thus users are confronted with a key difference between regular bits and qubits, and must rethink their error correction strategy accordingly. Here the in-game employee journal provides hints about a viable strategy for quantum error correction: that by entangling each qubit with ancillas prior to transmission, thereby `spreading' the information from a single qubit amongst a group of multiple qubits, the original qubit state can still be recovered even after some (limited) errors. Rather than working out the corrections required for each error syndrome mathematically, users can instead just run trial simulations in order to directly infer the errors corresponding to each syndrome. A solution to level H.C based on the bit-flip code is shown in Fig. \ref{fig:err}(b); although this passes the main level objective is does not meet the higher accuracy threshold required for the bonus star objective. Although, conceptually, it is relatively straight-forward to improve the accuracy of the bit-flip code by using more ancilla qubits, doing this within the limited space and processing time afforded by TQF remains a significant design challenge. An example of a more complicated solution that does achieve the bonus objective is given in Fig. \ref{fig:err}(c).

\begin{figure} [!t] 
\begin{center}
\includegraphics[width=8.5cm]{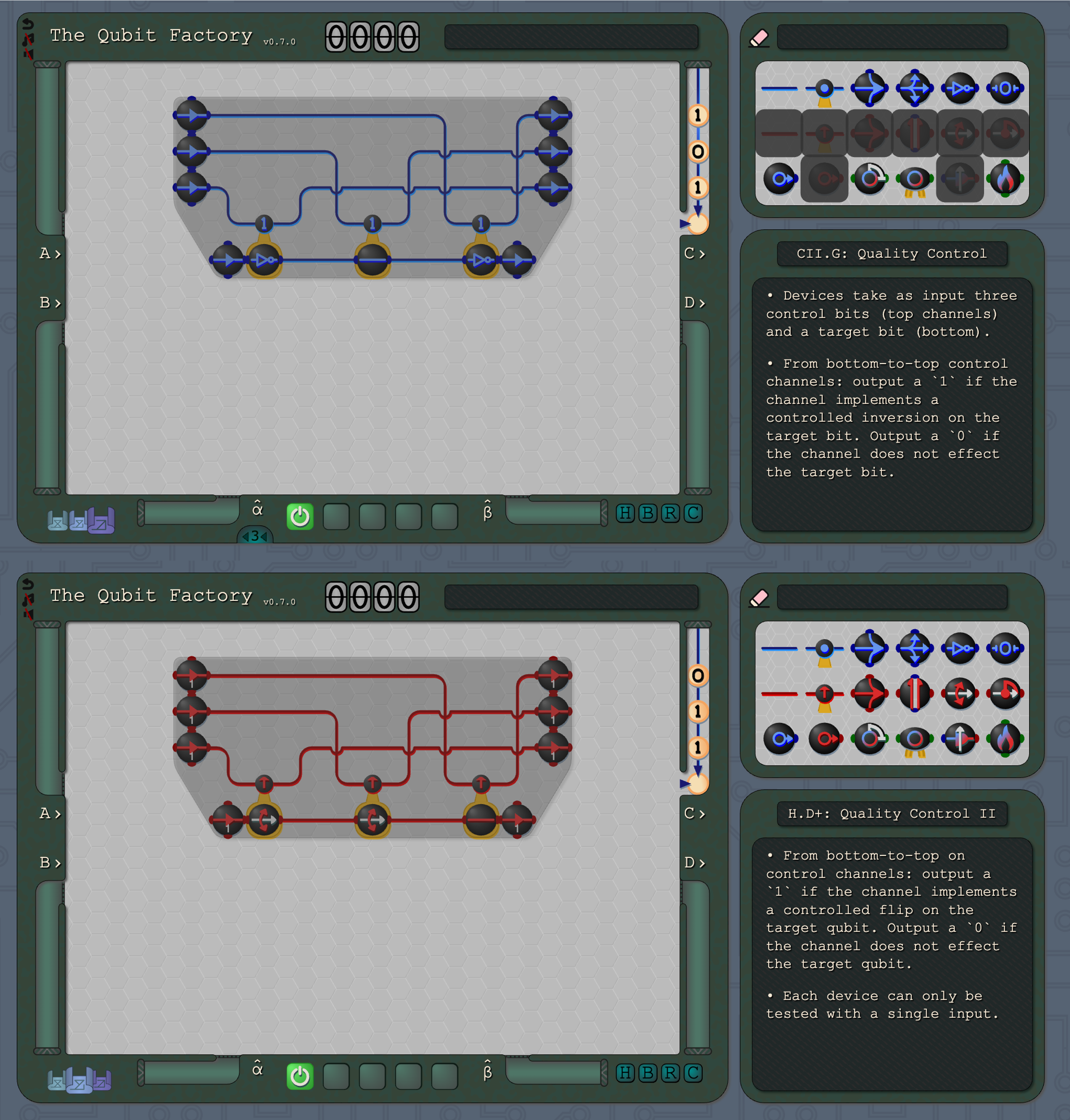}
\caption{Screenshots of levels CII.G (top) and H.D+ (bottom), which task the user with characterizing a set of classical or quantum devices respectively, with each device encoding a hidden binary function on three bits. The classical device requires at least three sets of inputs to be fully characterized, while the quantum device can be fully characterized with only a single set of inputs using the Bernstein-Vazirani algorithm\cite{Algo5}. Jointly these levels demonstrates to the user a situation where quantum computing has a distinct computational advantage over classical computing.}
\label{fig:device}
\end{center}
\end{figure}

A final example that we briefly remark upon is from levels CII.G and H.D+, which aim to illustrate a quantum advantage via the Bernstein–Vazirani algorithm\cite{Algo5}. Level CII.G, as shown in Fig. \ref{fig:device}, presents the user with a device that has three control channels, each of which may or may not individually implement a controlled inversion on a target channel (such that the device encodes a hidden binary function on three bits). Users are tasked with `testing' devices using appropriately chosen inputs in order to determine this hidden binary function. Upon investigation users will quickly find that at least three sets of inputs are needed to characterize each device, since only a single bit of information can be given from each set of inputs. At a much later stage of progress in the game users encounter level H.D+, as shown in Fig. \ref{fig:device}, which presents an identical task but with a quantum version of the device. In principle this task could be solved by re-using the same strategy employed in level CII.G for the classical devices; however an additional constraint is imposed in the quantum setting that only allows for a single set of inputs per device. It follows that in order to successfully complete the level it is necessary for the user to re-invent the Bernstein–Vazirani algorithm\cite{Algo5}, and thus discover for themselves an example where quantum computing has a demonstrable advantage over classical computation. While this may seem exorbitant expectation from any user not already knowledgeable of the Bernstein–Vazirani algorithm, several factors ease this burden: (i) the level immediately preceding H.D+ covers separately a key component of the Bernstein–Vazirani algorithm and (ii) users can easily experiment with different inputs, potentially allowing discovery through trial-and-error.

\subsection{Quantum computing content:} As demonstrated by the examples from the previous section, many levels in TQF have solutions that directly correspond to well-known algorithms and protocols from quantum computing. These include:
\begin{itemize}
  \item Superdense coding\cite{Algo6} (level H.A+).
  \item Quantum teleportation\cite{Algo3} (level H.B).
  \item Entanglement distillation\cite{Algo1,Algo2} (level H.B+).
  \item Classical/quantum error correction\cite{Error1,Error2} (levels CII.E, H.C+).
  \item The Bernstein–Vazirani algorithm\cite{Algo5} (levels CII.G, H.D+).
  \item Quantum repeaters\cite{Algo7} (level QIII.C).
  \item The CHSH game\cite{Algo8} (level QIII.F).
\end{itemize}
In order to facilitate ease of access to any of these levels (i.e. without having to unlock them via the normal gameplay progression) the GitHub repository of TQF\cite{TQF} provides save files that allow access to these levels and/or provide example solutions. Additionally, several other levels are designed to cover aspects of the following topics:
\begin{itemize}
\item   Reversibility/unitary of quantum computation (levels QI.A, QII.C, QII.D, QII.F).
\item   Measurement uncertainty (levels QI.C, QIII.A).
\item   Single-qubit quantum state tomography  (levels QI.D, QII.B, QIII.E).
\item   Multi-qubit quantum state tomography  (levels QII.H, QIII.G, H.E, H.E+, H.F, H.F+)
\end{itemize}

\section{Discussion: Quantum Gamification} \label{sect:Gamification}
In this manuscript we have examined some of the reasons why explaining quantum computation (or, more broadly, quantum mechanics) in a meaningful way to the layperson can be a significant challenge. One of the reasons speculated for this difficulty is that people lack an intuitive understanding of quantum behavior, as they are unlikely to have had first-hand experience in observing quantum phenomena. Gamification of quantum phenomena provides a novel way to remedy this deficit: by allowing users to interact with a simulated quantum world they can potentially build intuition via hands-on experience and experimentation\cite{GameTheory1, GameTheory2, GameTheory3, GameTheory4, GameTheory5}. However there remains many challenges in designing a quantum game, particularly with respect to representing abstract quantum phenomena (e.g. wavefunctions and entanglement) visually in a comprehensible manner. Another potential challenge comes with presenting new information and concepts in interesting ways to players (i.e. without relying on excessive textual or mathematical explanations). 

As detailed throughout this manuscript, TQF has employed many strategies aimed to remediate these difficulties in gamifying quantum mechanics. One strategy was to impose limitations on the scope of the physics, e.g. by restricting to real-valued wavefunctions, to simplify the user experience while still providing an accurate, although not complete, model of quantum computing. In terms of the visual representations within TQF, qubit states and qubit transformations are depicted geometrically (i.e. as arrows on the disk that can be rotated or flipped) in order to connect with the users existing geometric intuition, while gates and components have been designed such that their function can be inferred visually without pre-existing knowledge. Wherever possible TQF introduces new concepts via gameplay, relying on text-based explanations only as an optional supplement. 

Similar to other browser based interfaces for constructing quantum circuits\cite{GUI1, GUI2} or online tools\cite{Games8,NoCode1}, TQF provides no-code programming interface for quantum computing. Such interfaces could play an important role in on-boarding users without a formal quantum background into quantum computing: they allow users to easily explore scenarios and circuit configurations that would otherwise require a non-trivial mathematical analysis to evaluate. In particular, a key feature of TQF is to allow users to process a circuit step-by-step and to examine intermediate states that arise, as to facilitate their understanding of the circuit and allow them to iterate with new designs quickly and easily. By providing the opportunity for users to experiment with quantum circuits and algorithms in an approachable and user-friendly environment, TQF could provide a gateway to sophisticated quantum software frameworks\cite{OpenSource1, OpenSource2, OpenSource3, OpenSource4, OpenSource5, OpenSource6} and/or simulations on actual quantum hardware such as those mediated by Amazon Braket\cite{Braket1}. This could be especially useful for technically skilled engineers/scientists from other fields who are now entering the burgeoning field of quantum computing in some capacity.

In the future we plan to add additional levels and content to TQF (user suggestions are welcome in this regard) and may also consider adapting its structure such that it could be interfaced with existing quantum software frameworks.

\section{Acknowledgements} \label{sect:Acknowledgements}
Special thanks goes to Andrew Keller who, in addition to providing play-testing and content review, was instrumental in supporting bringing The Qubit Factory to release. Thanks to Alex Kubica, Ash Milsted and those within Amazon community that took part in the beta testing for providing feedback and suggestions. Thanks also go to John Preskill for giving permission to borrow his name (in a secret Easter egg).



\begin{thebibliography}{99}


\bibitem{TQF}
G. Evenbly, {\it The Qubit Factory}, \\Game: \href{https://www.qubitfactory.io}{https://www.qubitfactory.io}. \\Github: \href{https://github.com/awslabs/the-qubit-factory}{https://github.com/awslabs/the-qubit-factory}. \\Youtube: \href{https://www.youtube.com/@TheQubitFactory}{https://www.youtube.com/@TheQubitFactory}.


\bibitem{Nielsen}
M. A. Nielsen and I. L. Chuang, {\it Quantum Computation and Quantum Information}, Cambridge University Press, Cambridge; New York (2010).


\bibitem{Impact1}
J. P. Dowling and G. J. Milburn, {\it Quantum technology: the second quantum revolution}, Philos. Trans. R. Soc. Lond. Ser. Math. Phys. Eng. Sci. 361(1809), 1655–1674 (2003).

\bibitem{Impact2}
R. de Wolf, {\it The potential impact of quantum computers on society}, Ethics Inf. Technol. 19(4), 271–276 (2017).

\bibitem{Impact3}
J. Preskill, {\it Quantum computing in the NISQ era and beyond}, Quantum 2, 79 (2018).

\bibitem{Impact4}
J.-F. Bobier et al., {\it What happens when ‘if’ turns to ‘when’ in quantum computing?}, (2021).\\ \href{https://www.bcg.com/publications/2021/building-quantum-advantage}{https://www.bcg.com/publications/2021/building-quantum-advantage} (accessed 20 Dec 2023).

\bibitem{Impact5}
R. Waters, {\it Goldman Sachs predicts quantum computing 5 years away from use in markets}, Financial Times (2021). \href{https://www.ft.com/content/bbff5dfd-caa3-4481-a111-c79f0d38d486}{https://www.ft.com/content/bbff5dfd-caa3-4481-a111-c79f0d38d486} (accessed 20 Dec 2023).


\bibitem{Growth1}
E. Gibney, {\it Quantum gold rush: the private funding pouring into quantum start-ups}, Nature 574(7776), 22–24 (2019).

\bibitem{Growth2}
E. Hazan et al., {\it The next tech revolution: quantum computing}, (2020). \href{https://www.mckinsey.com/fr/our-insights/the-next-tech-revolution-quantum-computing}{https://www.mckinsey.com/fr/our-insights/the-next-tech-revolution-quantum-computing} (accessed 20 Dec 2023).


\bibitem{Workforce1}
P. E. Vermaas, {\it The societal impact of the emerging quantum technologies: a renewed urgency to make quantum theory understandable}, Ethics Inf. Technol. 19(4), 241–246 (2017).

\bibitem{Workforce2}
L. Nita et al., {\it The challenge and opportunities of quantum literacy for future education and transdisciplinary problem-solving}, Res. Sci. Technol. Educ., 1–17 (2021).

\bibitem{Workforce3}
C. Hughes et al., {\it Assessing the needs of the quantum industry}, IEEE Trans. Educ., 1–10 (2022).

\bibitem{Workforce4}
M. Kaurand, A .Venegas-Gomez, {\it Defining the quantum workforce landscape: a review of global quantum education initiatives}, Opt. Eng. 61(8), 081806 (2022).


\bibitem{Early1}
S. E. Economou, T. Rudolph, and E. Barnes, {\it Teaching quantum information science to high-school and early undergraduate students}, arXiv:2005.07874 (2020).

\bibitem{Early2}
A. Perry et al., {\it Quantum computing as a high school module}, arXiv:1905.00282 (2020).

\bibitem{Early3}
C. Foti et al., {\it Quantum physics literacy aimed at K12 and the general public}, Universe 7(4), 86 (2021).

\bibitem{Early4}
IEEE Quantum Education, {\it Report on how to support high school teachers interested in teaching and mentoring students in quantum science}.\\ \href{https://ed.quantum.ieee.org/teaching-high-school-discussion-report}{https://ed.quantum.ieee.org/teaching-high-school-discussion-report} (accessed 20 Dec 2023).


\bibitem{OnlineResource1}
J. R. Wootton et al., {\it Teaching quantum computing with an interactive textbook}, IEEE Int. Conf. Quantum Comput. and Eng. (QCE), pp. 385–391 (2021).

\bibitem{OnlineResource2}
A. Matuschak and M. Nielsen, {\it Quantum country}, (2019). \href{https://quantum.country}{https://quantum.country} (accessed 20 Dec 2023).


\bibitem{Games1}
A. Anupam, R. Gupta, A. Naeemi and N. Jafarinaimi, {\it Particle in a Box: An Experiential Environment for Learning Introductory Quantum Mechanics}, in IEEE Transactions on Education, vol. 61, no. 1, pp. 29-37, Feb. 2018.

\bibitem{Games2}
J. D. Weisz, M. Ashoori, and Z. Ashktorab, {\it Entanglion: a board game for teaching the principles of quantum computing}, in Proc. 2018 Annu. Symp. Comput.-Human Interaction in Play, CHI PLAY ’18, pp. 523–534, Association for Computing Machinery, New York, NY, USA (2018).

\bibitem{Games3}
C. Cantwell, {\it Quantum chess: developing a mathematical framework and design methodology for creating quantum games}, arXiv:1906.05836 (2019). Quantum Chess, \href{https://quantumchess.net/}{https://quantumchess.net/}.

\bibitem{Games4}
B. R. La Cour et al., {\it The virtual quantum optics laboratory}, arXiv:2105.07300 (2021).

\bibitem{Games5}
L. Nita et al., {\it Inclusive learning for quantum computing: supporting the aims of quantum literacy using the puzzle game quantum Odyssey}, arXiv:2106.07077 (2021).

\bibitem{Games6}
S. Zaman Ahmed et al., {\it Quantum composer: a programmable quantum visualization and simulation tool for education and research}, Am. J. Phys. 89(3), 307–316 (2021).

\bibitem{Games7}
Google Quantum AI, {\it The Qubit Game}, (2022). \href{https://quantumai.google/education/thequbitgame}{https://quantumai.google/education/thequbitgame}.

\bibitem{Games8}
P. Migdał et al., {\it Visualizing quantum mechanics in an interactive simulation - Virtual Lab
by Quantum Flytrap}, Opt. Eng. 61(8), 081808 (2022). 


\bibitem{GameTheory1}
S. Ornes, {\it Science and culture: quantum games aim to demystify heady science}, Proc. Natl. Acad. Sci. U. S. A. 115(8), 1667–1669 (2018).

\bibitem{GameTheory2}
A. Anupam et al., {\it Design challenges for science games: the case of a quantum mechanics game}, Int. J. Des. Learn. 11(1), 1–20 (2020).

\bibitem{GameTheory3}
M. Mykhailova and K. M. Svore, {\it Teaching quantum computing through a practical software-driven approach: experience report}, in Proc. 51st ACM Tech. Symp. Comput. Sci. Educ., Association for Computing Machinery, New York, pp. 1019–1025 (2020).

\bibitem{GameTheory4}
Z. C. Seskir et al., {\it Quantum games and interactive tools for quantum technologies outreach and education}, Opt. Eng. 61(8), 081809 (2022).

\bibitem{GameTheory5}
Quantum AI Foundation, {\it Quantum Games Hackathon}, \href{https://www.qaif.org/contests/quantum-games-hackathon}{https://www.qaif.org/contests/quantum-games-hackathon} (accessed 20 Dec 2023).


\bibitem{Engineer1}
C. G. Almudever et al., {\it The engineering challenges in quantum computing}, Design, Automation \& Test in Europe Conference \& Exhibition (DATE), 2017, Lausanne, Switzerland, 2017, pp. 836-845.\\ \href{https://doi.org/10.23919/DATE.2017.7927104}{https://doi.org/10.23919/DATE.2017.7927104}.

\bibitem{Engineer2}
{\it Quantum Computing: Progress and Prospects}. National Academies of Sciences, Engineering, and Medicine (2019). Washington, DC: The National Academies Press. \href{https://doi.org/10.17226/25196}{https://doi.org/10.17226/25196}.

\bibitem{Engineer3}
N. P. de Leon et al., {\it Materials challenges and opportunities for quantum computing hardware}. Science 372, abb2823 (2021). \href{https://doi.org/10.1126/science.abb2823}{https://doi.org/10.1126/science.abb2823}


\bibitem{Error1}
S. J. Devitt et al., {\it Quantum error correction for beginners}, Rep. Prog. Phys. 76, 076001 (2013).\\ \href{http://dx.doi.org/10.1088/0034-4885/76/7/076001}{http://dx.doi.org/10.1088/0034-4885/76/7/076001}

\bibitem{Error2}
D. A. Lidar and T. A. Brun, {\it Quantum Error Correction}, Cambridge Univ. Press (2013).


\bibitem{Zach1}
Z. Barth, {\it Zachtronics LLC}.\\ \href{https://www.zachtronics.com/}{https://www.zachtronics.com/}

\bibitem{Zach2}
{\it Human Resource Machine}, Tomorrow Corporation (2015). \\ \href{https://tomorrowcorporation.com/humanresourcemachine}{https://tomorrowcorporation.com/humanresourcemachine}

\bibitem{Zach3}
{\it Prime Mover}, 4Bit Games (2018). \\ \href{https://www.4bitgames.com/primemover}{https://www.4bitgames.com/primemover}


\bibitem{GUI1}
C. Gidney, {\it Quirk: Quantum Circuit Simulator}, (2019).\\ \href{https://github.com/Strilanc/Quirk}{https://github.com/Strilanc/Quirk}.

\bibitem{GUI2}
IBM Quantum Platform, {\it IBM Quantum Composer}.\\ \href{https://quantum.ibm.com/composer}{https://quantum.ibm.com/composer}.


\bibitem{Factorio1}
{\it Factorio}, Wube Software (2016).\\ \href{https://www.factorio.com/}{https://www.factorio.com/}


\bibitem{Bloch}
Technically, when describing real-valued wavefunctions on the Bloch sphere, one still needs azimuthal angles $\varphi=0$ and $\varphi=\pi$. We instead allow the `polar' angle to be in the range $\theta\in (-\pi,\pi]$, where negative angles substitute for cases with azimuthal angle $\varphi=\pi$. 


\bibitem{Algo1} 
C. H. Bennett, H. J. Bernstein, S. Popescu, and B. Schumacher, {\it Concentrating partial entanglement by local operations}, Phys. Rev. A 53, 2046 (1996).

\bibitem{Algo2} 
C. H. Bennett, G. Brassard, S. Popescu, B. Schumacher, J. A. Smolin, and W. K. Wootters, {\it Purification of Noisy Entanglement and Faithful Teleportation via Noisy Channels}, Phys. Rev. Lett. 76, 722 (1996).

\bibitem{Algo3} 
C. H. Bennett, G. Brassard, C. Crépeau, R. Jozsa, A. Peres, and W. K. Wootters, {\it Teleporting an unknown quantum state via dual classical and Einstein-Podolsky-Rosen channels}, Phys. Rev. Lett. 70, 1895 (1993).

\bibitem{Algo4} 
W. K. Wootters and W. H. Zurek, {\it A single quantum cannot be cloned}, Nature 299, 802–803 (1982).

\bibitem{Algo5} 
E. Bernstein and U. Vazirani, {\it Quantum Complexity Theory}, SIAM Journal on Computing 26 (5): 1411–1473 (1997).

\bibitem{Algo6} 
C. H. Bennett and S. J. Wiesner, {\it Communication via one- and two-particle operators on Einstein-Podolsky-Rosen states}, Phys. Rev. Lett. 69, 2881 (1992).

\bibitem{Algo7} 
K. Azuma, S. E. Economou, D. Elkouss, P. Hilaire, L. Jiang, H.-K. Lo, and I. Tzitrin, {\it Quantum repeaters: From quantum networks to the quantum internet}, Rev. Mod. Phys. 95, 045006 (2023).

\bibitem{Algo8} 
J. F. Clauser, M. A. Horne, A. Shimony, and R. A. Holt, {\it Proposed Experiment to Test Local Hidden-Variable Theories}, Phys. Rev. Lett. 23, 880 (1969).


\bibitem{NoCode1} 
H. Silvério et al., {\it Pulser: An open-source package for the design of pulse sequences in programmable neutral-atom arrays}, Quantum 6, 629 (2022). \href{https://doi.org/10.22331/q-2022-01-24-629}{https://doi.org/10.22331/q-2022-01-24-629}.


\bibitem{OpenSource1}
J. R. Johansson, P. D. Nation, and F. Nori, {\it QuTiP: an open-source Python framework for the dynamics of open quantum systems}, Comput. Phys. Commun. 183, 1760–1772 (2012).

\bibitem{OpenSource2}
M. Fingerhuth, T. Babej, and P. Wittek, {\it Open source software in quantum computing}, PLOS ONE 13, e0208561 (2018).

\bibitem{OpenSource3}
N. Killoran et al., {\it Strawberry fields: a software platform for photonic quantum computing}, Quantum 3, 129 (2019).

\bibitem{OpenSource4}
V. Bergholm et al., {\it PennyLane: automatic differentiation of hybrid quantum-classical computations}, \href{https://doi.org/10.48550/arXiv.1811.04968}{arXiv:1811.04968} (2020).

\bibitem{OpenSource5}
M. S. Anis et al., {\it Qiskit: an open-source framework for quantum computing}, (2021). \\ \href{https://github.com/Qiskit/qiskit}{https://github.com/Qiskit/qiskit}.

\bibitem{OpenSource6}
M. Fingerhuth, {\it Open-source quantum software projects}, (2022). \\ \href{https://github.com/qosf/awesome-quantum-software}{https://github.com/qosf/awesome-quantum-software}.


\bibitem{Braket1}
Amazon Web Services, {\it Amazon Braket}, 2020.\\ \href{https://aws.amazon.com/braket/}{https://aws.amazon.com/braket/}.

\end{thebibliography}
\end{document}